# Robust Electromagnetic Interference (EMI) Elimination via Simultaneous Sensing and Deep Learning Prediction for RF Shielding-free MRI

Yujiao Zhao[1,2], Linfang Xiao[1,2], Vick Lau[1,2],
Yilong Liu[1,2], Alex T. Leong[1,2], and Ed X. Wu[1,2*]

[1]Laboratory of Biomedical Imaging and Signal Processing
The University of Hong Kong, Hong Kong SAR, People's Republic of China
[2]Department of Electrical and Electronic Engineering
The University of Hong Kong, Hong Kong SAR, People's Republic of China

[*]Correspondence to:
    Ed X. Wu, Ph.D.
    Department of Electrical and Electronic Engineering
    The University of Hong Kong, Hong Kong SAR, China
    Tel: (852) 3917-7096
    Fax: (852) 3917-8738
    Email: ewu@eee.hku.hk

# ABSTRACT

At present, MRI scans are performed inside a fully-enclosed RF shielding room, posing stringent installation requirement and unnecessary patient discomfort. We aim to develop an electromagnetic interference (EMI) cancellation strategy for MRI with no or incomplete RF shielding. In this study, a simultaneous sensing and deep learning driven EMI cancellation strategy is presented to model, predict and remove EMI signals from acquired MRI signals. Specifically, during each MRI scan, separate EMI sensing coils placed in various spatial locations are utilized to simultaneously sample environmental and internal EMI signals within two windows (for both conventional MRI signal acquisition and EMI characterization acquisition). Then a CNN model is trained using the EMI characterization data to relate EMI signals detected by EMI sensing coils to EMI signals in MRI receive coil. This model is utilized to retrospectively predict and remove EMI signals components detected by MRI receive coil during the MRI signal acquisition window. We implemented and demonstrated this strategy for various EMI sources on a mobile ultra-low-field 0.055 T permanent magnet MRI scanner and a 1.5 T superconducting magnet MRI scanner with no or incomplete RF shielding. Our experimental results demonstrate that the method is highly effective and robust in predicting and removing various EMI sources from both external environments and internal scanner electronics at both 0.055 T (2.3 MHz) and 1.5 T (64 MHz), producing final image signal-to-noise ratios that are comparable to those obtained using a fully enclosed RF shielding. Our proposed strategy enables MRI operation with no or incomplete RF shielding, alleviating MRI installation and operational requirements. It is also potentially applicable to other scenarios of accurate RF signal detection or discrimination in presence of external and internal EMI or RF sources.

# INTRODUCTION

Magnetic resonance imaging (MRI) is intrinsically superior to other medical imaging modalities (e.g., computed tomography and positron emission tomography), because it is non-invasive, non-ionizing, inherently quantitative and multi-parametric[1,2]. However, conventional MRI scanners require specialized and expensive installations due to infrastructure requirements, e.g., site preparation to host the large magnets that typically weigh 3000-4500 kg, magnetic and radiofrequency (RF) shielding, electricity to drive power-consuming electronics, and water requirement for gradient cooling[3]. Consequently, the vast majority of clinical MRI scanners are housed inside fully enclosed RF shielding room on ground floors of large hospitals and clinics, severely hindering the accessibility and patient-friendliness of MRI in modern healthcare[4].

Recently, there has been a growing impetus to develop MRI scanners at ultra-low-field (ULF) strengths (i.e., below 0.1 Tesla or T)[4-13] for low-cost clinical imaging. Preliminary results demonstrated that these ULF MRI developments produced clinically valuable information for brain pathology diagnosis[10-15]. These ULF scanners also eliminate the need for a magnetic shielding cage because of dramatic fringe field reduction, yet many of them still require bulky and enclosed RF shielding to prevent external electromagnetic interference (EMI) signals during data acquisition. Such requirement precludes the portability of ULF MRI scanners for truly point-of-care applications (e.g., in intensive care units and surgical suites).

Several solutions have been recently developed to tackle the RF shielding cage requirement for ULF MRI[9,16-20]. One group used simple conductive cloth to cover the subject during scanning[9]. This passive method could alter and mitigate EMI from external environments, but its performance was suboptimal. Moreover, it was inadequate to deal with EMI from internal sources, such as scanner console, gradient/RF amplifiers, and power supplies. Alternatively, active methods were also proposed to reduce or eliminate external EMI. One group utilized magnetometers to sense environmental EMI and remove EMI signals in MRI receive coil via an adaptive suppression procedure[16]. This method was hardware demanding and only yielded limited success. An analytical approach was proposed to estimate EMI signals in MRI receive coil from EMI signals detected by EMI sensing coils using the frequency domain transfer functions among coils[17]. More recently, it was extended for time domain implementation as linear convolutions and with an adaptive procedure[18]. This method eliminated EMI substantially but could only produce very satisfactory brain imaging results when used together with conductive cloth and body surface electrode for EMI pickup.

The aforementioned active EMI elimination methods are based on a simple electromagnetic phenomenon, that is, the relationship of EMI signals detected by EMI sensing coils and MRI receive coil can be analytically characterized by the coupling or transfer functions among coils. However, in realistic unshielded imaging environments, EMI signals could be emitted by various sources from external environments as well as internal scanner electronics. Further, the EMI signals can change dynamically due to EMI sources and surrounding environments with various nature and behaviors. Such practical issues can complicate or degrade the performance of these analytical methods.

Intuitively, a deep learning driven method is preferable over the existing analytical approaches[16-20] for more robust EMI prediction and elimination. This is because neural networks potentially offer the ability of approximating the coupling relationships among coils from a subset of nonlinear functions, thus robustly predicting EMI signals detected by MRI receive coils especially in presence of complex external and internal EMI signals.

This study presents a novel simultaneous sensing and deep learning driven EMI cancellation strategy for RF shielding-free MRI[12,21,22]. It eliminates EMI signals from acquired MRI signals by establishing the relationships among EMI signals detected by EMI sensing coils and MRI receive coils via deep learning. This method works effectively and robustly with regards to various and dynamically varying EMI sources as demonstrated on our home-built mobile ULF 0.055 T head MRI scanner and a 1.5 T whole-body clinical MRI scanner with no or incomplete RF shielding.

# METHODS

## EMI Cancellation via Simultaneous Sensing and Deep Learning

An EMI cancellation strategy is presented to model, predict and remove EMI signals from acquired MRI signals by taking advantages of the well-established MRI multi-receiver electronics (previously developed for parallel imaging) and convolutional neural networks (CNNs), as illustrated in **Figure 1**. EMI sensing coils are strategically placed around scanner and inside electronic cabinet to actively detect radiative EMI signals from both external environments and internal scanner electronics (**Figure 1A**). Within each TR during the scanning, main MRI receive coil and EMI sensing coils simultaneously sample data within two acquisition windows, one is for the conventional MRI signal acquisition, the other is chosen for acquiring the EMI characterization data (no MRI signals due to no RF excitation, i.e., EMI signals only) (**Figure 1B**). After each scan, data sampled by both MRI receive coil and EMI sensing coils within the second window (i.e., EMI characterization window) are used to train a CNN model that can relate the 1D temporal EMI signal received by multiple EMI sensing coils to the 1D temporal signals received by MRI receive coil for each frequency encoding (FE) signal or k-space line (**Figure 1C**). This model is then applied to predict the EMI signal component in MRI receive coil signal for each FE line within MRI signal acquisition window based on the EMI signals simultaneously detected by EMI sensing coils. Subsequently, the predicted EMI signal component is subtracted or removed from the MRI receive coil signals. This procedure is repeated for all individual FE lines, creating EMI-free k-space data prior to any averaging or/and image reconstruction.

## Evaluation on a Mobile 0.055 T Head MRI Scanner

The proposed strategy was implemented and evaluated on our home-built mobile ULF 0.055 T MRI scanner[12] with no RF shielding. All experiments involving human subjects were approved by the local institutional board and written information consents were obtained.

Phantom and brain datasets were acquired with 1 MRI receiving coil and 10 EMI sensing coils. These EMI sensing coils were deployed around and inside the scanner to detect EMI signals that were from both external environment and those generated internally by gradient/RF electronics during MRI scanning. Each of them was fabricated by wounding copper wire on a 3D printed coil holder (diameter = 5 cm) and was tuned to the Larmor frequency (2.32 MHz). The detected EMI signal was passed through a two-stage preamplifier module (first-stage: Gain = 30 dB; second stage: Gain =30 dB, for input Vpp < 60 mV). The placement of these EMI sensing coils is depicted in **Figure 1A**. Three were placed in the vicinity of the patient head holder, two on each side (left and right) underneath the patient bed, and two in the vicinity of gradient and RF amplifiers inside the electronic cabinet, and one underneath the scanner.

The effectiveness of the proposed EMI cancellation strategy was quantitatively evaluated by comparing two scenarios, where a fully enclosed RF shielding cage was installed or removed, respectively (as shown in **Figure 2**). Further, to examine the robustness of our method in presence of various and dynamically varying EMI sources, phantom imaging experiments were conducted with artificially-generated EMI: (i) additional broadband EMI generated from a nearby source and (ii) additional swept frequency EMI generated from a nearby source (center frequency = 2.32 MHz, sweep span = 100 kHz, frequency points = 101, and continuous sweeping cycle period = 4 s).

3D GRE and FSE sequences were implemented and optimized for the 0.055 T head MRI scanner to demonstrate the RF shielding-free brain MRI. T1W datasets were acquired with 3D GRE (TR/TE = 52/13 ms, flip angle = 40°, FOV = 250×250×320 mm$^3$, acquisition slice thickness = 10 mm, acquisition matrix = 128×126×32, and NEX = 2). T2W and FLAIR datasets were acquired using 3D FSE with FOV = 250×250×320 mm$^3$ and slice thickness = 10 mm (T2W: TR/TE = 1500/202 ms, ETL = 21, acquisition matrix = 128×126×32, and NEX = 2; FLAIR: TR/TE = 500/129 ms, ETL = 13, acquisition matrix = 128×117×32, and NEX = 4). Elliptic 2D phase encoding (PE) patterns were used by the 3D GRE and FSE sequences to reduce total scan time. For above three protocols, their scan times were approximately 5.5 mins, 7.5 mins, and 7.5 mins, respectively.

### Evaluation on a 1.5 T Whole-body MRI Scanner

To demonstrate the applicability of our proposed strategy to high-field MRI, we also applied our proposed method to remove EMI signals in phantom k-space data that were acquired on a 1.5 T MRI scanner. Specifically, phantom datasets were acquired on a whole-body 1.5 T clinical MRI scanner using a 16-channel coil with RF shielding room door closed or open. Four defective RF channels were treated as EMI sensing coils because they received little MR signals but strong EMI signals, thus acting as EMI sensing coils. 2D GRE protocol with TR/TE = 420/9 ms, BW = 25 kHz, acquisition matrix = 200×200×20, and NEX = 2 was used.

### Model Implementation

For each scan, datasets sampled by MRI receive coil and EMI sensing coils within the EMI characterization window (i.e., contains EMI signals only) were used to train a five-layer CNN model. Note that each layer within the CNN model was a combination of 2D convolution, batch normalization and rectified linear unit (ReLU), except the last layer where only convolution operation was performed. The kernel sizes of the five convolutional layers were 11×11, 9×9, 5×5, 1×1, and 7×7, respectively, with the corresponding number of channels being 128, 64, 32, 32, and 2. The complex data were processed by feeding the real and imaginary parts into the network as two separate input channels. The input of the network was a 3D matrix with a size of $N_{FE} \times N_s \times 2$, where $N_{FE}$, $N_s$ and 2 are number of points in one FE line, number of EMI sensing coils, and number of channels corresponding to real and imaginary parts of the raw data. The output of the network was a 2D matrix with a size of $N_{FE} \times 2$. During the training stage, the mean squared error loss (i.e., L2 loss) was minimized using Adam optimizer[23] with $\beta 1 = 0.9$, $\beta 2 = 0.999$, and initial learning rate = 0.0005. The CNN model was implemented with a batch size of 16 for 20 epochs on a Quadro RTX 8000 GPU and Intel Core i9-10900X CPU. The trained model was then applied to remove EMI signals from acquired MRI signals using datasets sampled by both MRI receive coil and EMI sensing coils within the MRI signal acquisition window (i.e., contains MRI + EMI signals).

Our proposed deep learning driven EMI cancellation approach was implemented using PyTorch 1.8.1

package on Ubuntu 18.04.5 LTS (Linux 5.4.0-77-generic). All source code can be obtained online (https://github.com/joey024/Deep-Learning-Driven-EMI-Cancellation) or from the authors upon request.

## RESULTS

**Figure 3** compares 0.055 T human brain imaging in two scenarios, where a fully enclosed RF shielding cage was installed or removed (as illustrated in **Figure 2**). Two brain imaging experiments (with and without RF shielding cage) were conducted from the same normal adult subject using 3D FSE FLAIR protocol with NEX = 2. Note that a small stable narrowband EMI noise was still present in the images before EMI cancellation when RF shielding cage was installed (due to MRI console EMI leakage, as pointed by green arrows). It indicated that passive EMI cancellation method (e.g., using RF shielding cage) was inadequate to prevent internal EMI from scanner electronics. When the RF shielding cage was removed, the brain images were severely contaminated by the internal narrowband EMI, as well as narrowband and broadband EMI from external environments. However, with the proposed deep learning EMI cancellation strategy, both internal and external EMI were effectively removed. The effectiveness of the proposed EMI cancellation method was quantitatively examined. Here, image noise levels were calculated from the difference images between the first and second individual complex images[24]. Without the deep learning EMI elimination procedure, the average noise level without RF shielding cage was significantly higher (1.423 in standard deviation, SD) than that with RF shielding cage (0.581) as expected. After deep learning EMI elimination, the average noise level without RF shielding cage substantially decreased to 0.588 (from 1.423), which was only 1.2% higher than that obtained with RF shielding cage before EMI elimination (0.581) and 4.8% higher than that after EMI elimination (0.561). The results demonstrated that, in absence of the RF shielding cage, the proposed method was able to provide nearly complete EMI removal, producing final image noise levels as low as those obtained with a fully enclosed RF shielding cage installed (within 5% range) in human brain imaging experiments.

**Figure 4** present the EMI cancellation results of 0.055 T human brain imaging with no RF shielding cage for three typical clinical contrasts (i.e., T1W, T2W, and FLAIR). Before EMI cancellation, the EMI related image noise or artifacts exhibited different patterns in images with different contrasts. This was mainly because the three protocols were implemented with different FE directions (indicated by white arrows). With the proposed approach, the EMI signal components in the MRI receive coil signals were accurately predicted and effectively removed, resulting in substantially improved image quality.

The proposed EMI cancellation results for 0.055 T phantom imaging with no RF shielding cage are shown in **Figure 5**. Phantom datasets were acquired using 3D FSE T2W protocol and with artificially-generated external EMI: (i) additional broadband EMI and (ii) additional swept frequency EMI. Before EMI cancellation, one can easily discern and identify numerous sources of external (environmental or/and artificially generated) EMI and internal (e.g., from console) EMI, as indicated by yellow arrows. Moreover, these EMI signals were indeed not static across the duration of each scan. For example, it can be clearly seen from the frequency spectra (i.e., Fourier transform of FE line, or FT of FE) that spectral characteristics of some EMI sources changed dynamically, either slowly or very rapidly, over time during scanning. Nevertheless, the spectral and image results after EMI cancellation, clearly demonstrated the effectiveness and robustness of the proposed method in the presence of diverse and dynamically varying EMI signals.

**Figure 6** illustrates the tolerance of our proposed method to EMI signal change arises from a change of subject body position during scanning. Human brain images were acquired using 4-average 3D FSE FLAIR protocol. Four individual images sequentially acquired during the FLAIR protocol are shown. Before EMI removal, it can be seen that a moderate body position change in the middle of a scan (i.e., bending of lower legs between 2nd and 3rd acquisitions) caused changes in detected EMI characteristics (in terms of both frequency and magnitude). This demonstrated another dynamically varying EMI scenario that can be encountered in realistic imaging setting due to subject position changes, or movements of nearby attending staff and equipment during scan. With the proposed simultaneous sensing and deep learning driven EMI elimination strategy, all EMI signal components were effectively removed, indicating its robustness in practice.

**Figure 7** shows the results from a 1.5 T clinical MRI sited within a fully enclosed RF shielding room. Note that, even with the RF shielding room door closed, EMI artifacts were still presented in the images (due to imperfection of the front-end electronics inside the scanner room). With shielding room door open, more EMI artifacts are observed due to external EMI sources as expected. The proposed strategy eliminated both internal and external EMI (arising from hardware imperfections and incomplete RF shielding), and significantly improved the image quality when door was open, leading to image SNRs generally better than those with RF shielding room door closed. These results clearly indicated the applicability of our proposed EMI sensing and cancellation method to high-field MRI, i.e., at high frequency regime.

## DISCUSSION

In this study, an EMI cancellation strategy is presented to model, predict and remove EMI signals from acquired MRI signals, thus eliminating the need for RF shielding. This method produces significant and robust EMI removal without introducing any detriment to MRI signals. Further, the resulting image SNRs are highly comparable to those obtained using a fully enclosed RF shielding cage, indicating a nearly complete EMI removal of the proposed method in RF shielding-free brain MRI at 0.055 T (~2.3 MHz). The effectiveness of our proposed strategy is also successfully demonstrated in clinical MRI at 1.5 T (~64 MHz).

The prediction of EMI signals is based on an electromagnetic phenomenon, that is, the properties of RF signal propagation among any radiative (e.g., air) or/and conductive media (e.g., surrounding EMI emitting structures such as power lines, MRI hardware pieces and cables, and imaging object or human body) are fully dictated by the electromagnetic coupling among these media or structures[25,26]. As revealed in recent studies[16-20], such coupling relationships are able to be analytically characterized by frequency domain coupling or transfer function among structures. However, in realistic unshielded imaging environments, the nature of EMI signals is diverse and complex because EMI signals from various sources (e.g., external environments and internal low-cost scanner electronics) can change dynamically during a scan. In this study, such practical issues are resolved by simultaneous EMI sensing and deep learning driven EMI prediction and elimination. This is because the nature of deep learning is supposed to enable a versatile model that robustly establishes the relationships among EMI signals detected by EMI sensing coils and MRI receive coil. Further, unlike most data-driven deep learning methods, the proposed method does not require a large training dataset from historical scans. Instead, it directly derives a neural network model from scan-specific EMI characterization data, such that it can be more robust to the dynamically varying EMI.

The above hypothesis has been well supported by the 0.055 T imaging results. Both phantom and human brain imaging (of a cohort of ~100 normal volunteers and patients) have demonstrated that the proposed EMI cancellation strategy worked effectively and robustly for external and internal EMI from various sources, even when the EMI sources and their spectral characteristics changed dynamically during MRI scan (**Figures 3** to **5**). Moreover, EMI signals received by MRI receive coil can also be influenced by the human body, which serves as an effective antenna[27,28] for EMI reception in a shielding-free MRI setting. Varying body size or weight can alter the level and characteristics of EMI signal pickup by body and subsequently detected by MRI receive coil. Human body position change during MRI scan can also alter the EMI signals detected by MRI receive coil due to the change in electromagnetic coupling between human body and surrounding EMI emitting sources. Importantly, this study has illustrated that, with minor body position change, the proposed method could still effectively remove undesirable EMI signals, yielding substantially improved image quality (**Figure 6**).

It is noteworthy to emphasize the potentially broad applicability of the proposed EMI cancellation strategy. First, the present study demonstrated that our strategy works robustly not only on a RF shielding-free mobile 0.055 T MRI at ~2.3 MHz, but also on a clinical 1.5 T MRI that operated at much higher RF frequency (~64 MHz). Second, our proposed strategy is highly applicable to various RF signal detection applications. Future studies can evaluate such EMI cancellation scheme for other non-MRI RF signal detection applications (e.g., radio astronomy or radar), where we anticipate further improvements of using the proposed EMI cancellation approach than conventional EMI mitigation methods[29-31] when near or/and far EMI characteristics or/and environments become more complex.

# CONCLUSION

This study presents a simultaneous sensing and deep learning driven EMI cancellation strategy to model, predict and remove EMI signals from acquired MRI signals. It utilizes the well-established MRI multi-receiver electronics as well as the nature of neural network, leading to a versatile model that can robustly establish the relationships among EMI signals detected by EMI sensing coils and MRI receive coil.

More importantly, the proposed strategy of active EMI sensing and deep learning prediction can offer a novel and effective approach for various RF signal detection/discrimination applications in presence of complex external and internal EMI or RF sources.

# ACKNOWLEDGEMENTS

This study is supported in part by Hong Kong Research Grant Council (R7003-19, HKU17112120 and HKU17127121 to E.X.W. and HKU17103819, HKU17104020 and HKU17127021 to A.T.L.), and Lam Woo Foundation.

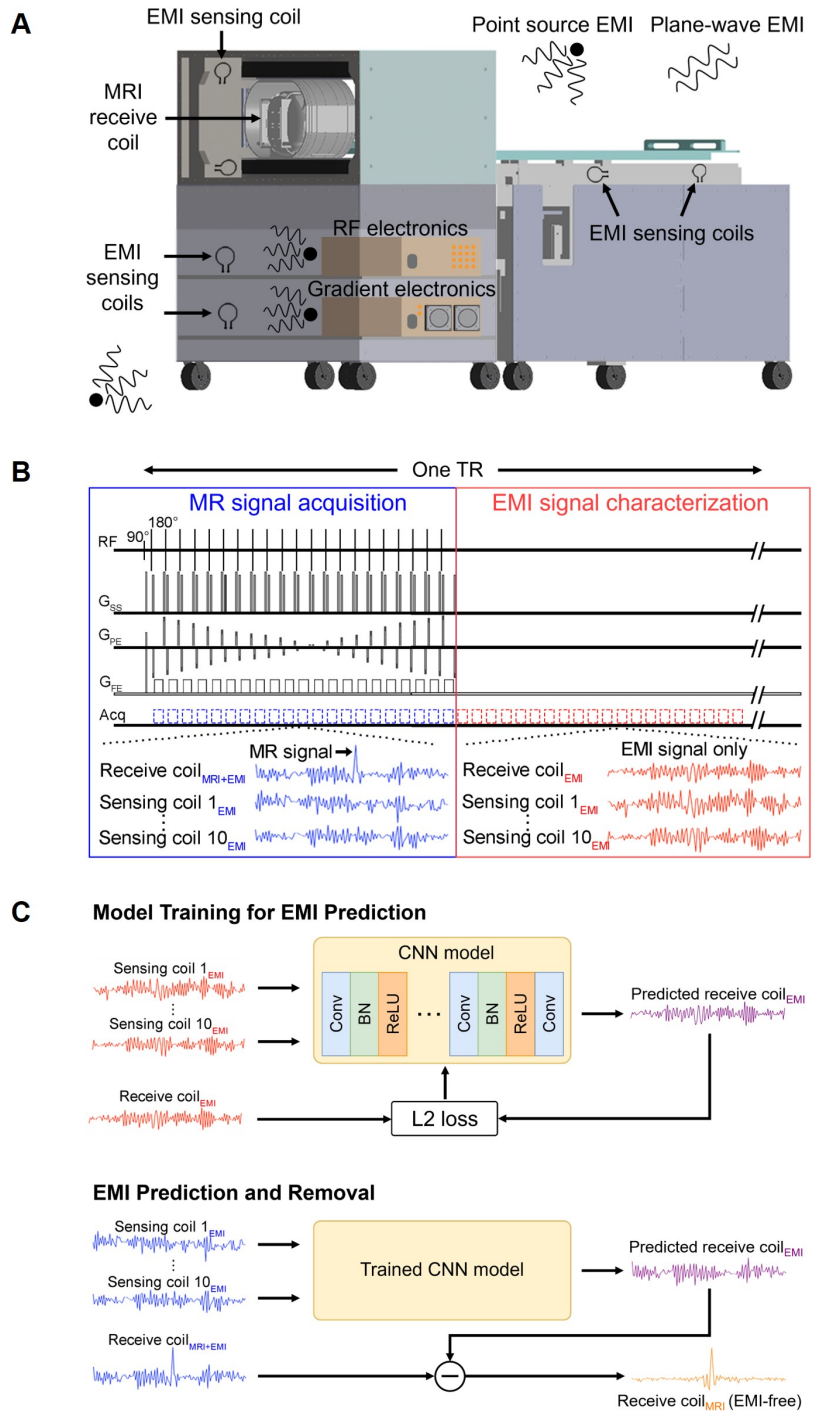

**Figure 1.** Simultaneous sensing and deep learning driven electromagnetic interference (EMI) prediction and elimination on a mobile ultra-low-field (ULF) 0.055 T permanent magnet brain MRI scanner with no RF shielding, with 2.32 MHz resonance frequency. (**A**) Multiple EMI sensing coils are placed around and inside scanner to actively detect EMI signals during MRI scan. (**B**) Illustration of 3D FSE acquisition windows for MRI signal collection and EMI signal characterization within two windows. (**C**) A CNN model is trained to establish the relationship between EMI signals received by MRI receive coil and sensing coils within the second window, i.e., EMI signal characterization window. The trained model is then utilized to predict EMI signal component detected by the MRI receive coil within the first window, i.e., the conventional MRI signal acquisition window.

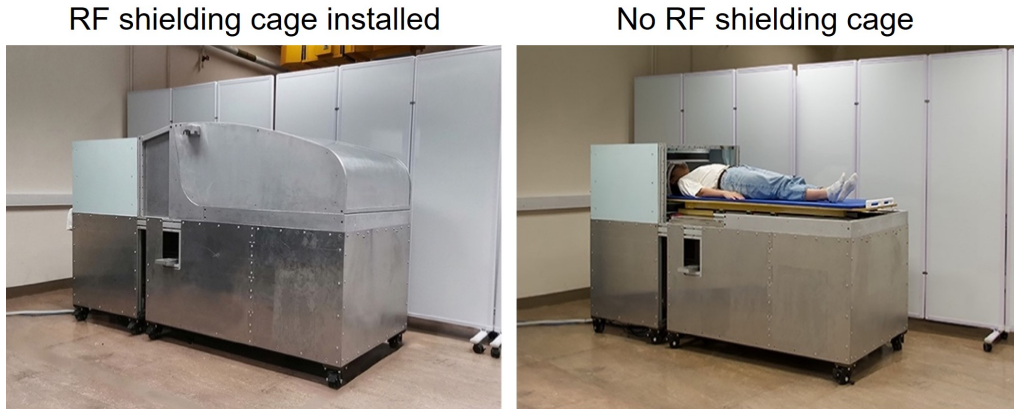

**Figure 2**. Experimental setups for quantitatively assessing the effectiveness of the proposed EMI cancellation method in eliminating EMI related image noise for 0.055 T mobile MRI scanner. For comparison, a custom-made RF shielding cage (consisting of aluminum covers) was installed to fully enclose the subject, or removed.

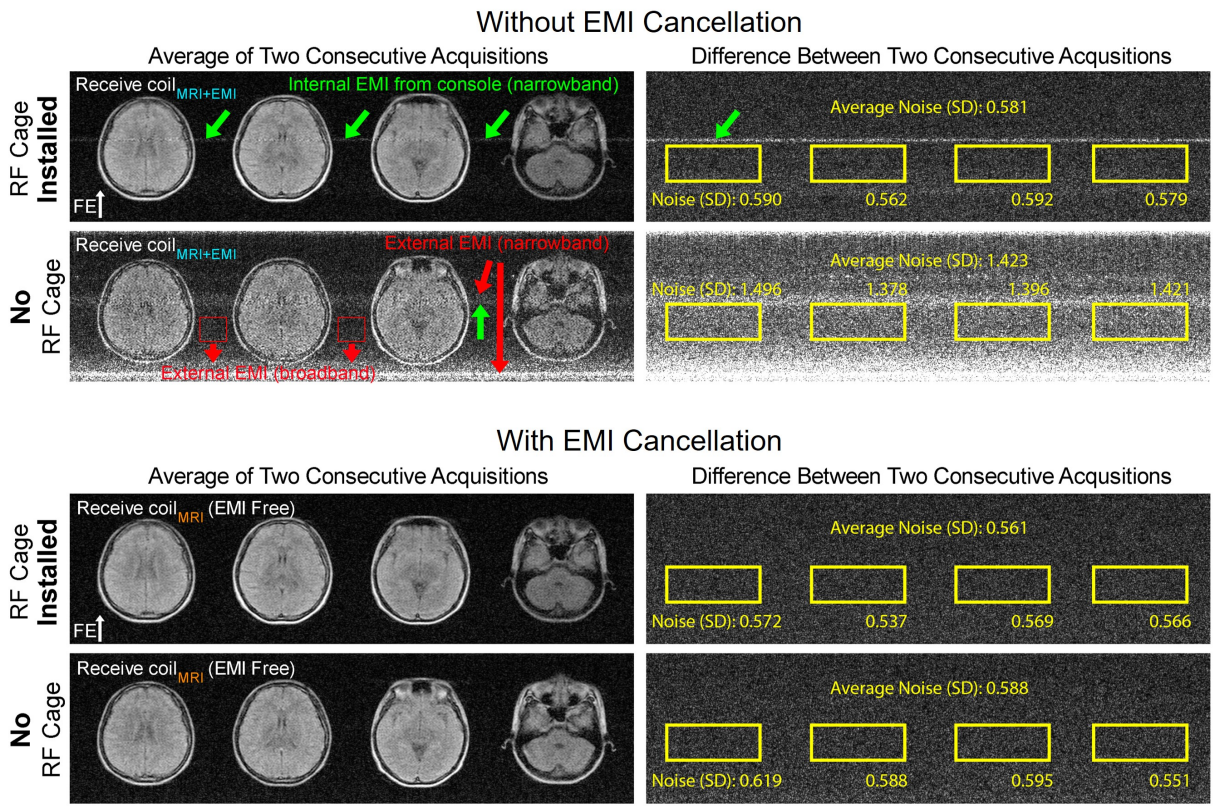

**Figure 3.** Deep learning driven EMI cancellation for 0.055T brain imaging. FLAIR images and corresponding noise images (**A**) with and (**B**) without EMI cancellation are shown. The images acquired with and without RF shielding were compared. Before EMI cancellation, an internal EMI signal (green arrows) was still present even when the RF shielding cage was installed. With no RF shielding, both external and internal EMI signals were present. The proposed method effectively eliminated external and internal EMI signals, leading to final image noise levels as low as those obtained with a fully enclosed RF shielding cage installed (within 5% range) in human brain imaging experiments.

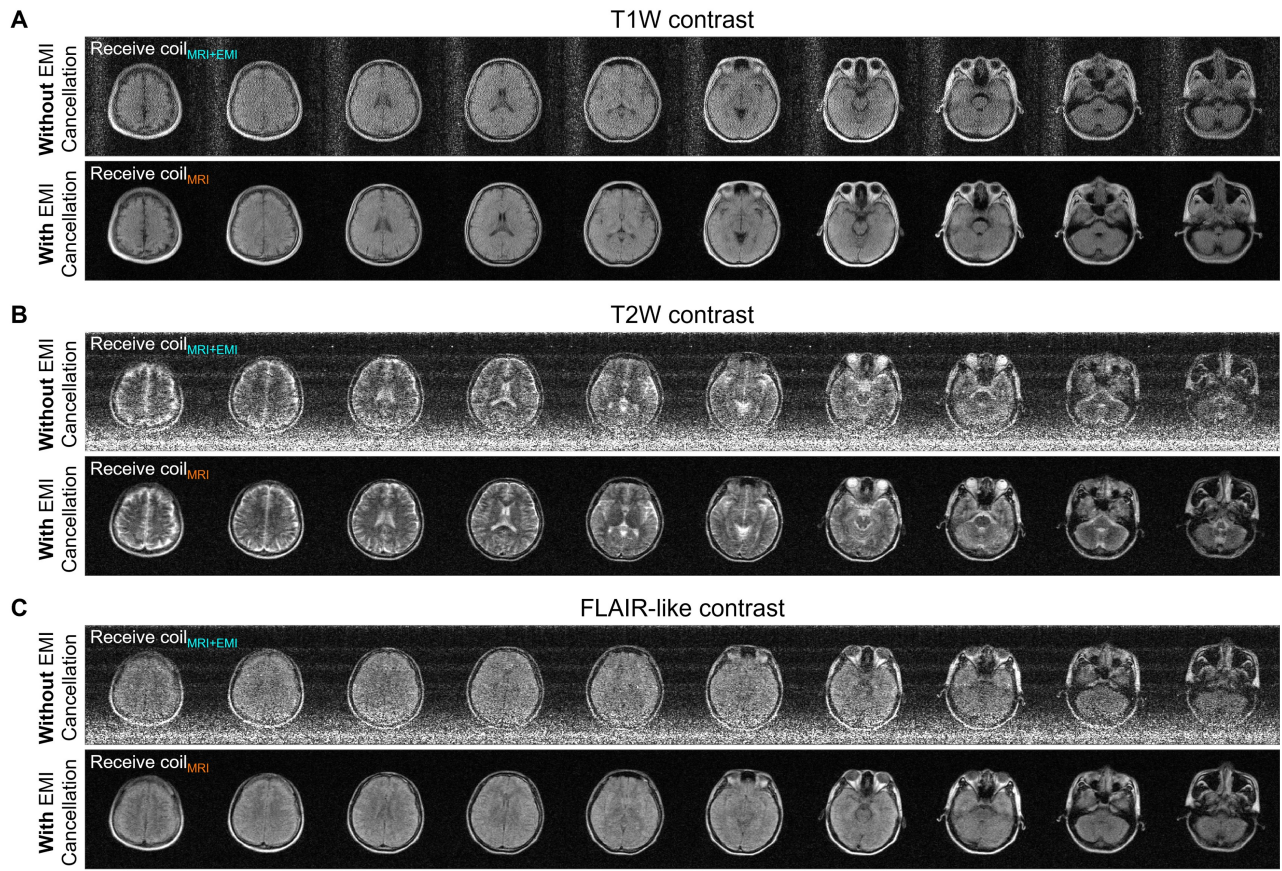

**Figure 4.** Deep learning driven EMI cancellation for 0.055 T human brain imaging with no RF shielding. Brain datasets were acquired from a healthy adult subject at (**A**) T1W, (**B**) T2W and (**C**) FLAIR-like contrasts. T1W, T2W and FLAIR-like datasets were acquired using 3D GRE, 3D FSE and 3D FSE with short TR, respectively. All images are displayed at a spatial resolution of 1×1×5 mm$^3$, while the acquisition resolution is approximately 2×2×10 mm$^3$. The proposed deep learning EMI cancellation strategy robustly predicted the EMI signals detected by the MRI receive coil, enabling MRI scan without any RF cage or shield.

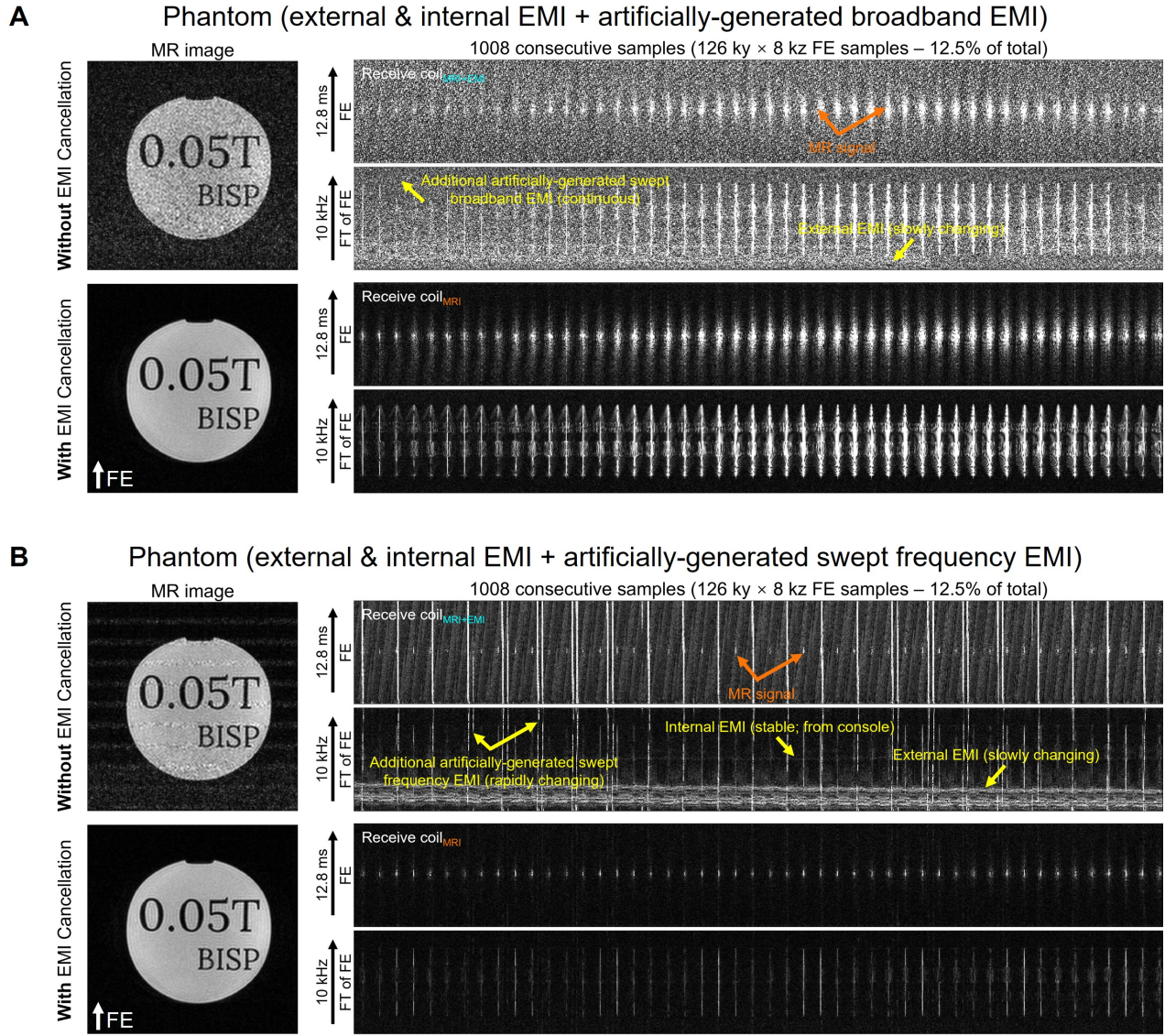

**Figure 5.** Deep learning driven EMI cancellation for 0.055 T phantom imaging with no RF shielding in presence of various external and internal EMI sources. T2W phantom datasets were acquired using 3D FSE and with artificially-generated EMI: (**A**) additional broadband EMI generated from a nearby source and (**B**) additional swept frequency EMI generated from a nearby source. MR images (left) and spectral analyses (right) of raw acquired data are shown. The results with and without EMI cancellation are displayed using the same brightness. The proposed EMI cancellation strategy eliminated all EMI signals from various sources, as directly demonstrated by the spectral and image results (i.e., the presence of MRI signals only and complete absence of any discernable EMI signals).

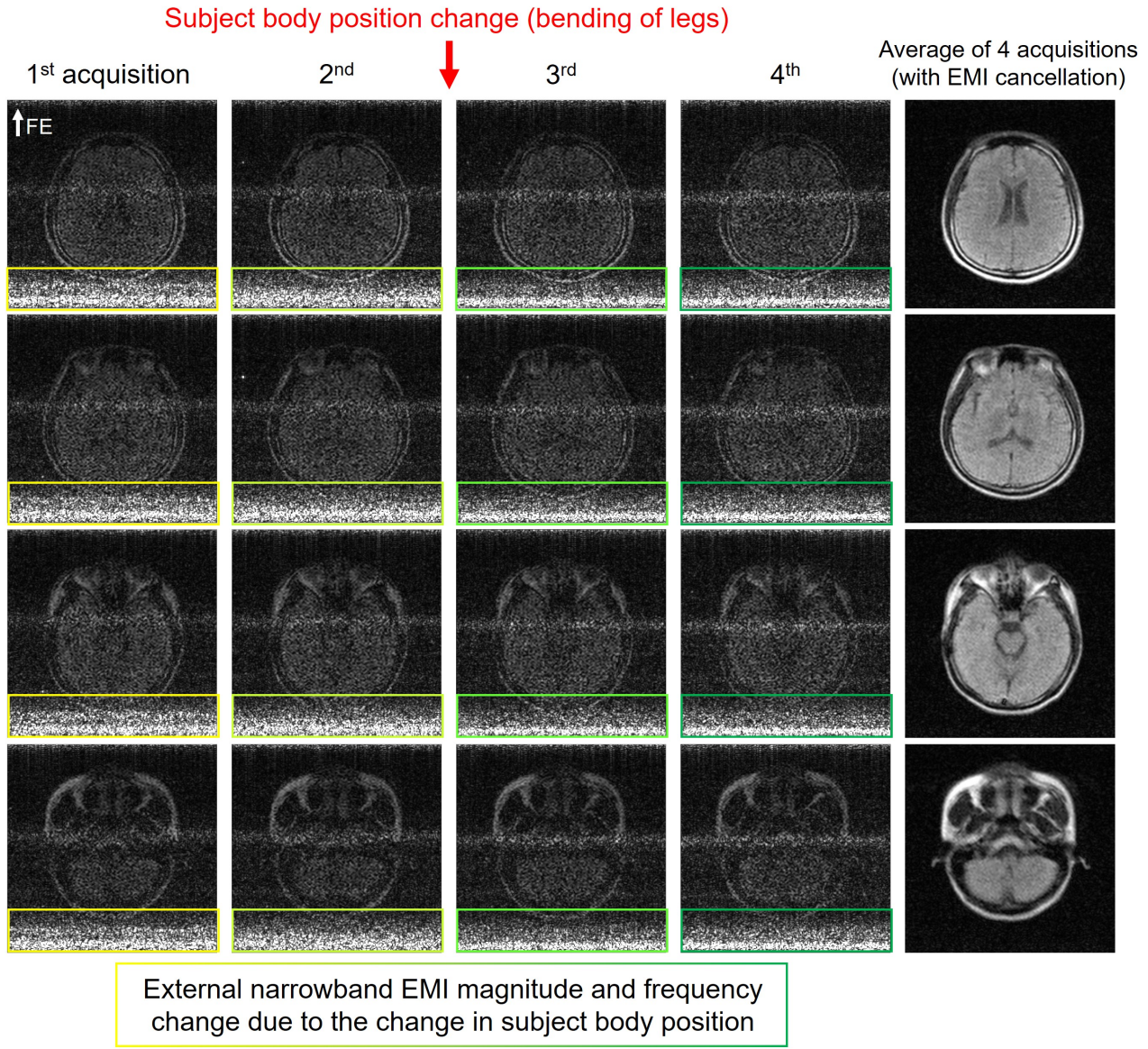

**Figure 6.** Changes in detected EMI signals due to a change of subject body position during a 4-average FLAIR scan. Four individual 3D FSE FLAIR image datasets were sequentially acquired before averaging from a normal adult. The corresponding images without EMI removal are shown on the left. Only four brain slices are shown in the four rows here. FE was along vertical direction. Before EMI cancellation, external narrowband EMI could be seen as the horizontal noise bands located at the bottom of these individual images (as well as the thinner horizontal noise bands around the middle of the images). Yellow and green boxes indicate the EMI characteristics before and after the body position change, respectively. It can be seen that a moderate subject body position change, i.e., bending of legs between the 2nd and 3rd acquisitions, led to notable difference in the external narrowband EMI frequency location and magnitude. With the proposed deep learning driven EMI elimination strategy, the EMI related artifacts were effectively removed as demonstrated by the final EMI-free images on the right.

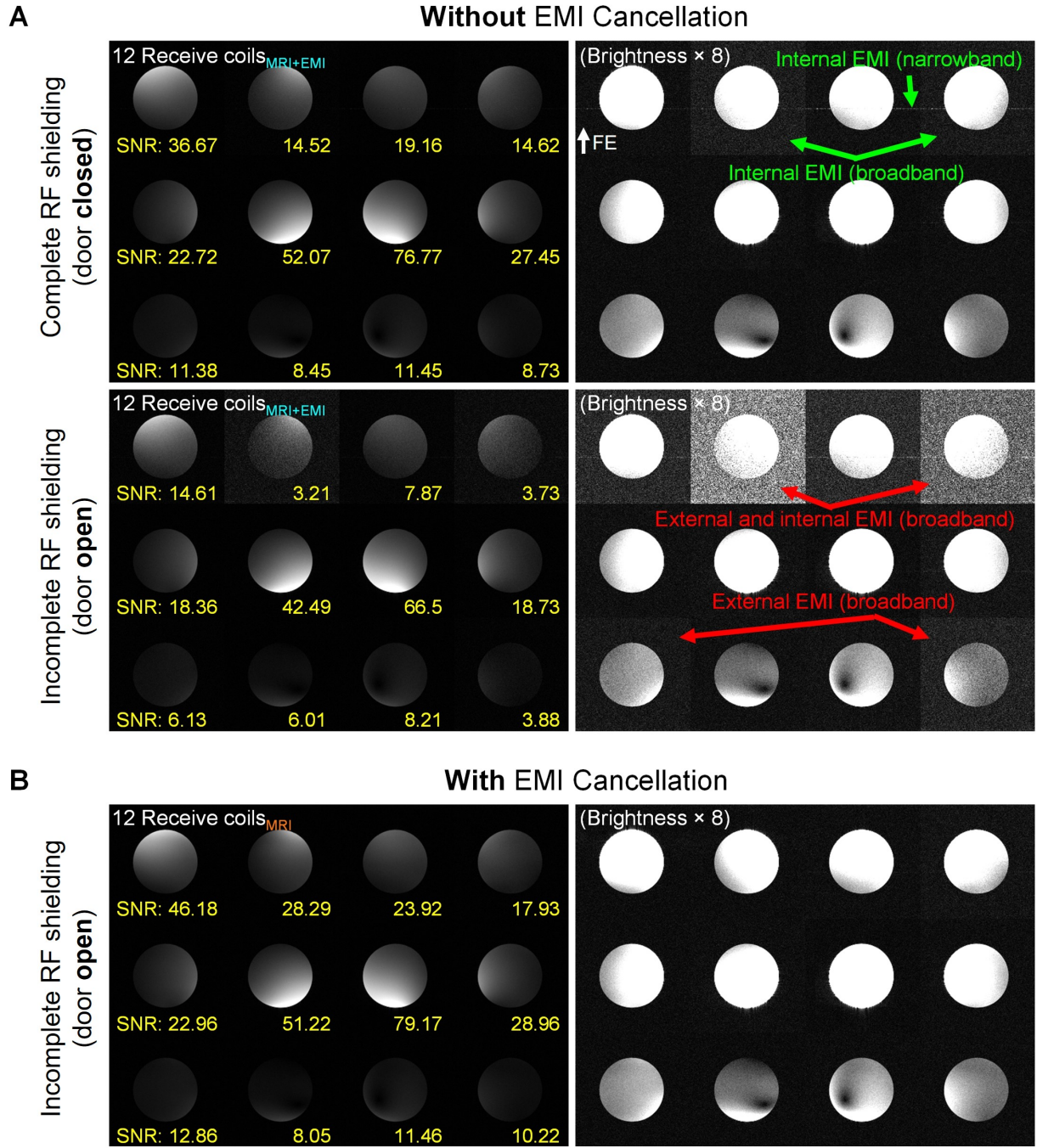

**Figure 7.** With incomplete RF shielding (i.e., RF shielding room door open) on a clinical 1.5 T superconducting magnet whole-body MRI scanner with 64 MHz resonance frequency, the method produced final phantom images with SNRs generally better than those when RF shielding room door was closed. Images (**A**) with and (**B**) without the EMI cancellation are shown. Before EMI cancellation, both narrowband and broadband internal EMI signals (generated by electronics inside scanner room) were present even when door was closed. The proposed strategy eliminated both external and internal EMI, and significantly improved the image quality when door was open. The results clearly indicated the applicability of our proposed EMI sensing and cancellation method to high-field MRI, i.e., at high frequency regime.